\documentclass[twocolumn,prd,amsmath,amssymb, aps,superscriptaddress, nofootinbib]{revtex4-1}
\usepackage[version=4]{mhchem} 
\usepackage[normalem]{ulem}
\usepackage{amsmath}
\usepackage{amssymb}
\usepackage{amsfonts}
\usepackage{mathrsfs}
\usepackage{graphicx}
\usepackage{xcolor}
\usepackage{xfrac}
\usepackage{footmisc}
\usepackage{pifont}
\usepackage{times}
\usepackage{epstopdf}
\usepackage{enumerate}
\usepackage[hidelinks]{hyperref}
\usepackage{enumitem}
\usepackage{slashed}
\usepackage{relsize}
\usepackage{epsfig}
\usepackage{verbatim}
\usepackage{bm}
\usepackage[utf8]{inputenc}
\usepackage{graphics}
\usepackage{subfigure}
\usepackage[english]{babel}
\usepackage{orcidlink}

\definecolor{zima_blue}{HTML}{1393C1}
\hypersetup{setpagesize=false,
bookmarksnumbered=true,
bookmarksopen=true, 
colorlinks=true,
linkcolor=zima_blue,
urlcolor=zima_blue,
citecolor=zima_blue,
linktocpage=false}

\begin{document}

\title{Dark matter production accompanied by gravitational wave signals during cosmological phase transitions
}

\author{Shuocheng Xu}
\email{xushuocheng@s.ytu.edu.cn}
\affiliation{Department of Physics, Yantai University, Yantai 264005, China}

\author{Ruiyu Zhou} 
\thanks{Corresponding author}
\email{zhoury@cqupt.edu.cn}
\affiliation{School of Science, Chongqing University of Posts and Telecommunications, Chongqing 400065, China}
\affiliation{Department of Physics and Chongqing Key Laboratory for Strongly Coupled Physics, Chongqing University, Chongqing 401331, China}

\author{Wei Cheng~\orcidlink{0000-0001-9943-4519}} 
\email{chengwei@cqupt.edu.cn}
\affiliation{School of Science, Chongqing University of Posts and Telecommunications, Chongqing 400065, China}
\affiliation{Department of Physics and Chongqing Key Laboratory for Strongly Coupled Physics, Chongqing University, Chongqing 401331, China}

\author{Xuewen Liu~\orcidlink{0000-0003-3652-7237}}
\thanks{Corresponding author}
\email{xuewenliu@ytu.edu.cn}
\affiliation{Department of Physics, Yantai University, Yantai 264005, China}

\begin{abstract}
 
We investigate the temperature-dependent production of feebly interacting massive  dark matter particle (FIMP DM) within a $Z_2$ model, incorporating two $Z_2$-odd scalar fields. In specific parameter regions, three distinct mechanisms individually dominate the production of the FIMP DM. These mechanisms include the semi-production process, commonly known as the ``exponential growth'' process, the three-body decay process, and the production from pair annihilations of the bath particles.
It is crucial to consider the thermal history during the evolution of FIMPs, as it involves multiple phase transitions occurring prior to the freeze-in of dark matter.
Consequently, the scalar masses experience thermal variations, leading to distinctive evolutionary trajectories for FIMPs when compared to scenarios without accounting for the thermal effects.
Notably, the unique patterns of FIMP evolution are accompanied by the production of gravitational waves, presenting promising opportunities for detection using forthcoming interferometers.

\end{abstract}

\maketitle

\section{Introduction}
\label{sec-intro}

Extensive evidence \cite{Zwicky:1933gu,Ostriker:1973uit,Peacock:2001gs,Bertone:2016nfn,Planck:2018vyg}  overwhelmingly supports the existence of dark matter (DM) as the dominant matter component of the Universe. However, despite decades of dedicated experimental efforts, the elusive nature of DM remains unresolved. Deciphering the fundamental properties of DM stands as one of the most pressing and fundamental challenges in contemporary particle physics. 
The Weakly Interacting Massive Particle (WIMP) has emerged as a leading candidate in addressing the DM problem, attracting extensive investigation \cite{Gunn:1978gr, Hut:1977zn, Lee:1977ua, Bertone:2004pz}. However, the parameter space of WIMP DM has faced substantial constraints imposed by direct detection experiments \cite{XENON:2018voc, CMS:2016lcl, MAGIC:2016xys, Arcadi:2017kky, PandaX-II:2016vec, LUX:2016ggv}, necessitating exploration of alternative DM production mechanisms. Among these mechanisms, the freeze-in mechanism has garnered significant interest as a viable alternative\cite{McDonald:2001vt,Choi:2005vq,Kusenko:2006rh,Hall:2009bx,Cheung:2011nn,Elahi:2014fsa,Arcadi:2015ffa,Bernal:2017kxu,Benakli:2017whb,Bernal:2018qlk,Bernal:2019mhf,Covi:2020pch,Khan:2020pso,Garcia:2020hyo,Bernal:2020qyu}.

The freeze-in mechanism, in contrast to the well-established freeze-out mechanism, gives rise to Feebly Interacting Massive Particle (FIMP) DM. FIMPs possess a small coupling to the visible sector, making them challenging targets for direct detection experiments and constraints. Note that the requirement of a small coupling for FIMP DM finds a natural explanation within clockwork models, providing a compelling theoretical framework for their production \cite{Kim:2017mtc, Kim:2018xsp, Goudelis:2018xqi, Choi:2015fiu, Kaplan:2015fuy, Giudice:2016yja}.
Among the most extensively studied production processes are the $1\to 2$ decays of thermal bath particles into FIMPs and the $2\to 2$ pair annihilation processes \cite{Bernal:2017kxu}, which have been thoroughly reviewed in Ref. \cite{Bernal:2017kxu}.
Recently, a novel production mechanism based on freeze-in through semi-production has been proposed \cite{Bringmann:2021tjr}. In this mechanism, DM particles are produced from the thermal bath via the transformation of heat bath particles, namely the inverse semi-annihilation process. For a small initial abundance of DM, this leads to an exponential growth in the DM number density.
Several studies have been conducted to explore this topic \cite{Hryczuk:2021qtz, Costa:2022oaa, Bhatia:2023yux, Liu:2023zah}.


However, it is of utmost importance to acknowledge the critical influence of finite temperature effects on the evolution of FIMP abundance. 
The thermal loop corrections contribute to the modification of particle masses, implying that the kinematics of DM production becomes temperature-dependent. Consequently, certain production channels may be accessible during specific cosmological epochs but kinematically closed during others. Another effect is the temperature dependence of the effective scalar potential, leading to variations not only in scalar masses but also in their vacuum expectation values (VEVs) in the early Universe.
Of particular significance is the scenario wherein the scalar potential demonstrates multiple disjoint minima and undergoes phase transitions between them. This phenomenon gives rise to a phase-dependent freeze-in mechanism for FIMPs. 
The impact of thermal effects in the early Universe on FIMP abundance has been previously examined in studies such as \cite{Baker:2017zwx, Bian:2018bxr, Bian:2018mkl, Deng:2020dnf, Bian:2021dmp,Azatov:2021ifm,Azatov:2022tii,Elor:2021swj, Shibuya:2022xkj}.

In this work, we investigate the temperature-dependent production of scalar FIMP DM within a new physics model featuring $Z_2$ symmetry. The model consists of two $Z_2$-odd scalar fields, $\Phi$ and $S$, with interaction terms $\lambda_1 \Phi S^3$ and $\lambda_2 \Phi^3 S$. We assign the role of FIMP DM to the field $S$, while $\Phi$ acts as the portal between the Standard Model (SM) sector and the dark sector. Additionally, $\Phi$ could also serve as a WIMP-like particle, evolving as a thermal relic. However, the abundance of $\Phi$ at present is negligible, leaving FIMP DM as the sole DM candidate in our scenario.
In this model, three different mechanisms or processes can dominate the production of FIMPs. These mechanisms include the semi-production process, denoted as $\Phi S \to SS$, the three-body decay process ($\Phi\to SSS$), and the production from pair annihilations of the bath particles ($\Phi\Phi\to SS$ and $\Phi\Phi\to \Phi S$).
To ensure a comprehensive analysis, we study each mechanism individually with specific benchmark parameters where they dominate the production.
Furthermore, we take into account the thermal history for the evolution of FIMPs, considering multiple phase transitions that occur prior to the freeze-in of DM. Incorporating the $\Phi$ field can induce a strongly first-order phase transition (FOPT) with a two-step pattern \cite{Cheng:2018ajh,Cheng:2018axr,Zhou:2018zli,Chiang:2020yym,Chiang:2019oms,Chao:2017vrq,Ramazanov:2021eya}. 
We would like to mention that the model under investigation in this study shares a similarity with the scalar potential presented in Ref. \cite{Costa:2022oaa}. However, it is imperative to highlight that the consideration of finite temperature effects on the production of FIMPs has not been accounted for.
The impact of finite temperature effects, specially on the semi-production and three-body decay processes, has not been explored in the literature. We find that this leads to a distinct evolutionary trajectory for FIMPs.
Moreover, the presence of a strong FOPT gives rise to the production of stochastic gravitational waves (GWs). Leveraging the $\Phi$-assisted two-step phase transition, the signal strength of the GWs is expected to be sufficiently high to be detectable by upcoming interferometers. 
This presents a complementary approach for investigating DM properties.

The rest of this paper is structured as follows.  
In Sec. \ref{sec:model}, we start with a general description of $Z_2$ model, incorporating two new scalar fields. Moving forward, Sec. \ref{sec:PT} focuses on the analysis of a two-step pattern of phase transition, where we examine the thermal masses of $S$ and $\Phi$. Subsequently, in Sec. \ref{sec:FIMP}, we delve into a thorough investigation of the production of FIMP DM. The study of GW signals is conducted in Sec. \ref{sec:GWs}. Finally, our findings and conclusions are summarized in Sec. \ref{sec:conclusion}.

\section{A  $Z_2$ model}
\label{sec:model}

Let us consider an extension of the SM by two real scalar singlets $S$ and $\Phi$, both charged under a new $Z_2$ symmetry. $S$ and $\Phi$ are assumed to transform, respectively, as $S\to-S$ and $\Phi\to-\Phi$, whereas SM fields are singlets of the $Z_2$. The Lagrangian, symmetric under $SU(3)\times SU(2)\times U(1)\times Z_2$, is given by,
\begin{align}\label{eq:potential}
 -{\cal L}\supset & ~ \mu_{H}^2|H|^2 + \mu_{\Phi}^2 \Phi^2 + \mu_{S}^2S^2 + \lambda_{H}|H|^4 + \lambda_{\Phi} \Phi^4 \nonumber \\ 
 &+\lambda_{S} S^4+ \lambda_{SH}S^2|H|^2
  + \lambda_{H\Phi} |H|^2 \Phi^2 \nonumber\\ 
 & + \lambda_{S\Phi}S^2\Phi^2+ \lambda_1 \Phi S^3 + \lambda_2 \Phi^3 S, 
\end{align}
where $H$ is the SM Higgs and $H=(0,(h+v_h)/\sqrt{2})^T$ in unitary gauge after getting VEV. The couplings $\lambda_{1,2}$ are assumed to be real. 
In our scenario, we consider the scalar field $S$ as a FIMP DM candidate. On the other hand, the scalar field $\Phi$ serves as the portal connecting the dark sector and the SM sector. Additionally, $\Phi$ can be considered as a WIMP-like DM candidate.
The field $\Phi$ has the potential to be in thermal equilibrium with the SM thermal bath through its coupling with the SM Higgs field. As a result, $\Phi$ can act as the parent particle responsible for the production of FIMP DM.
Various processes such as semi-production, pair annihilations of $\Phi$ particles, and three-body decays of $\Phi$, can be realized through interactions involving the scalar fields. 
To incorporate the strong constraint on the Higgs invisible decay, we intentionally suppress the coupling term between the Higgs boson and DM by setting $\lambda_{SH}=0$. The scalar field $\Phi$ does not acquire a VEV after EW symmetry breaking. 
Also, since our focus is on the freeze-in mechanism, we assume that the relevant couplings are much smaller than unity ($\ll 1$). This choice allows the model parameters to be less constrained by experimental limitations \cite{Hall:2009bx}.

The inclusion of additional scalar fields $\Phi$ beyond the SM scalar sector also introduces non-trivial dynamics into the evolution of the vacuum state. This can potentially give rise to a FOPT in the early universe. In this work, our focus is on a two-step phase transition pattern, which leads to a diverse DM production history.

\section{A two-step phase transition pattern}
\label{sec:PT}

The scalar field $\Phi$ may not acquire a VEV at zero temperature; however, it still plays a significant role in contributing to various types of phase transitions. Specifically, during the early stages of the universe, $\Phi$ has the capability to temporarily break the $Z_2$ symmetry with a nonzero VEV. As the universe cools down, the VEV of $\Phi$ eventually vanishes, leading to the restoration of the $Z_2$ symmetry.
This behavior exemplifies a typical two-step phase transition process. 
Moreover, the inclusion of the $\Phi$ field can induce a strongly FOPT, effectively preserving the baryon asymmetry from being washed out.

To account for the thermal corrections to the tree-level potential, we employ the high-temperature expansion approach \cite{Patel:2011th,Chao:2017vrq,Liu:2022jdq}. 
In this method, the effective finite temperature potential is given by
\begin{align}
V_T(h,\Phi,T) =& \frac{\mu_H^2 
+c_h(T)}{2} h^2 + \frac{\lambda_H}{4} h^4 + \frac{\lambda_{H\Phi}}{2} h^2 \Phi^2 \nonumber \\
&+ \frac{2\mu_\Phi^2 + c_\Phi(T)}{2} \Phi^2  
+ \lambda_\Phi \Phi^4, 
\end{align}
where
\begin{align}
&c_h(T) = \frac{1}{48}(9g^2 + 3g^{\prime2} + 12 y_t^2 + 24 \lambda_H + 4 \lambda_{H\Phi}) T^2,  \\
&c_\Phi(T) = \frac{1}{6} (2\lambda_{H \Phi} + \lambda_{S\Phi} + 6 \lambda_\Phi) T^2.
\end{align}
So the VEVs of Higgs $h$ and $\Phi$ 
in their each direction possess temperature dependence,  which can be written as,
\begin{equation}
v_h (T) = \pm \sqrt{\frac{-\mu_H^2 - c_h(T)}{ \lambda_H}}, 
\label{eqn:vevh}
\end{equation}
\begin{equation}
v_\Phi (T) = \pm \sqrt{\frac{-2\mu_\Phi^2 - c_\Phi(T)}{4 \lambda_\Phi}}. 
\label{eqn:vevphi}
\end{equation}
The thermal loop corrections and the temperature-dependent VEVs lead to variations in the particles' masses, which can be expressed formally as follows, 
\begin{align}
m_{S}^2 &= 2\mu_S^2 + 2\lambda_{S\Phi}v_{\Phi}(T)^2+\frac{1}{6}T^2(6\lambda_S+\lambda_{S\Phi}), 
\label{eqn:ms-t}
\end{align}
\begin{align}
m_{\Phi}^2 &= 2\mu_\Phi^2 + \lambda_{H\Phi}v_{h}(T)^2+12\lambda_{\Phi}v_\Phi(T)^2
+ c_\Phi (T), 
\label{eqn:mphi-t}
\end{align}
\begin{align}
m_{h}^2 &=\mu_H^2+3\lambda_H v_h(T)^2+\lambda_{H\Phi}v_\Phi(T)^2 +c_h(T). 
\label{eqn:mh-t}
\end{align}
The precise thermal masses in each phase will be determined by employing the method in the Appendix \ref{apdx:thermalmass} \footnote{
Besides the scalar mass corrections, scale running during thermal evolution may alter the coupling constants, potentially impacting phase transitions. However, given the minimal scale variation in the temperature range of the phase transitions in this work, as indicated in prior studies (e.g. \cite{Cai:2017tmh}), the impact of running couplings is negligible.}.

There are three essential temperatures characterizing the FOPT: the critical temperature $T_c$,  the bubble nucleation temperature $T_n$ and the percolation temperature $T_p$. 
$T_n$ is characterized by being lower than the critical temperature $T_c$, and it can be determined by performing the following computation \cite{Moreno:1998bq}, 
\begin{align}
\int_{t_c}^{t_n} {\frac{\Gamma(t)}{H(t)^3}} dt =  \int_{T_n}^{T_c} {\frac{\Gamma(T)}{H(T)^4 T}} dT =1,
\end{align}
where $\Gamma(T)$ is the bubble nucleation rate\cite{Affleck:1980ac,Linde:1981zj,Linde:1980tt}, which is defined by $\Gamma(T) \simeq A(T) e^{(-S_3/T)} $. $S_3$ is the bounce action for an $O(3)$-symmetric bounce solution:  
\begin{align}
S_3 = 4 \pi \int_{0}^{\infty} dr r^2 \left[ \frac{1}{2} \frac{d \phi_b}{dr}^2 + V_T(\phi_b,T)  \right], 
\end{align}
with $\phi_b = \Phi , h$ in this study. 
With the finite-temperature effective potential, the bounce action can be obtained by solving the following equation of motion:  
\begin{align}
\frac{d^2 \phi_b}{ dr^2} + \frac{2}{r} \frac{d \phi_b}{dr} - \frac{\partial V(\phi_b)}{\partial \phi_b} =0,  
\end{align}
with the boundary conditions being
\begin{align}
\frac{d\phi_b}{dr}\bigg|_{r=0} = 0, \; \; \; \mathop {\lim }\limits_{r \to \infty } \phi_b =\phi_b^{\rm false}.
\end{align}

After the nucleation of bubbles, another significant temperature is the percolation temperature, wherein the probability of encountering a point that remains in the false vacuum state is 0.7~\cite{Guth:1981uk,Ellis:2018mja}:
\begin{align}
&P(T_p) = \exp(-I(T_p)), \\
&I(T)=\frac{4\pi v_w^3}{3}  \int_{T}^{T_c} \frac{d T^{\prime} \Gamma(T^{\prime})}{H(T^{\prime})T^{\prime 4}} \left(\int_{T}^{T^{\prime}} \frac{d T^{\prime\prime}}{H(T^{\prime\prime})}\right)^3, 
\end{align}
where $v_w$ is the bubble wall velocity.

In order to carry out the necessary calculations, we utilize the package CosmoTransitions \cite{Wainwright:2011kj}. Our analysis involves the selection of three benchmark points  (BPs) that achieve strong FOPTs. These specific BPs are presented in Tab.~\ref{tab:PT-benchmarks}.

\begin{table}[htbp]
    \centering
    \begin{tabular}{|c|c|c|c|c|c|c|c|c|c|c|} 
  \hline
  BP & $m^0_\Phi ({\rm GeV})$ &  $\lambda_{H\Phi}$ &  $\lambda_{\Phi}$   &  $T_p ({\rm GeV})$& $ \frac{v_h(T_p)}{T_p}$& $\alpha$&  $\frac{\beta}{H_*}$ \\ 
  \hline
   BP1  &$128.8 $ & $0.491$ & $0.202$ & $63.1$ & 3.431& 0.17 & 282.2\\
   \hline
   BP2 &$131.3$ & $0.663$ & $0.482$  & $45.8$ &5.034& 0.46 & 121.6\\
   \hline
   BP3 &$80.0$ & $0.502$ & $0.410$ &  $45.1$ & 5.134&  0.33 & 1620.1\\
   
  \hline
 \end{tabular}  
    \caption{Benchmarks for strong FOPT. The couplings satisfy theoretical constraints, including vacuum stability and perturbative unitarity \cite{Kannike:2012pe}. $m^0_\Phi $ represents the zero-temperature mass of $\Phi$. Parameters $\alpha$ and  ${\beta}/{H_*}$ are essential for the GW productions, which will be discussed in Sec. \ref{sec:GWs}. }
    \label{tab:PT-benchmarks}
\end{table}

\begin{figure}[htbp]
\centering
\includegraphics[width=.465\textwidth]{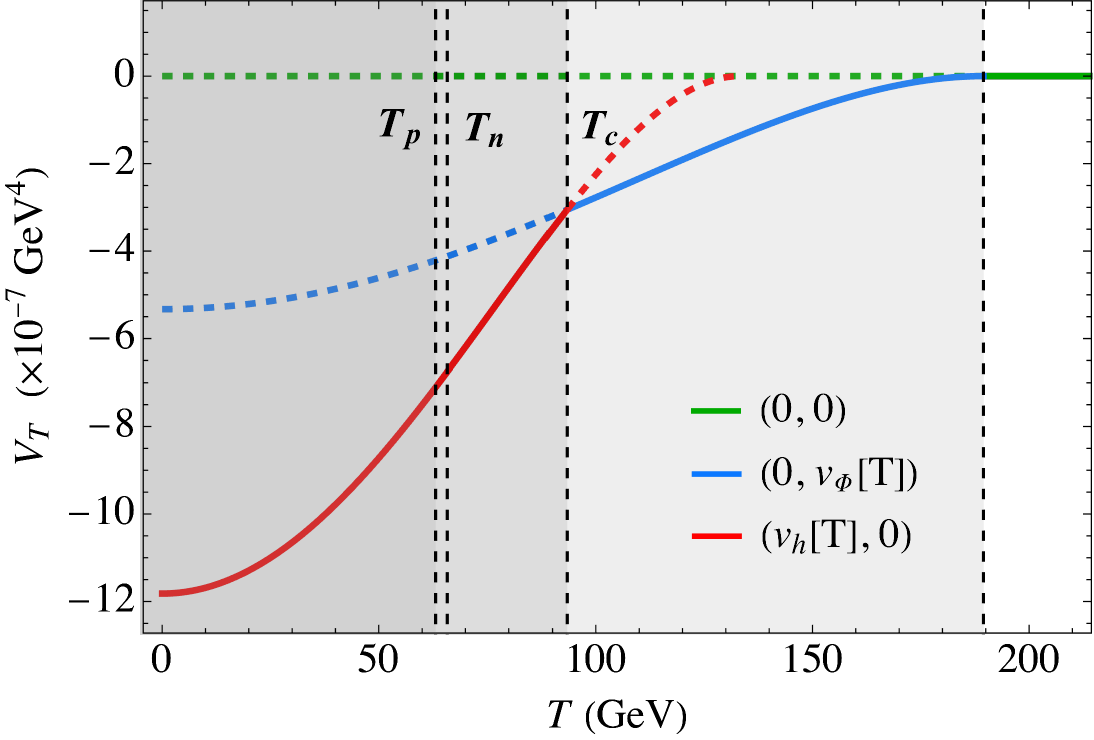}
\caption{Evolution of the effective potential minimum as function of temperature. The green, blue, and red lines correspond to the  potential minima at the symmetric phase, the $Z_2$ broken phase, and the EW broken phase, respectively. The model parameters are taken from BP1.}
\label{fig:potential_t}
\end{figure}

Fig.~\ref{fig:potential_t} displays the temperature-dependent evolution of the minimum of the effective potential, as demonstrated using the BP1 parameter set outlined in Tab.~\ref{tab:PT-benchmarks}. The minimum of $V_{\rm T}$ is obtained by substituting the fields with the VEVs determined from Eq. (\ref{eqn:vevh}) and Eq. (\ref{eqn:vevphi}). 

In the symmetric phase characterized by higher temperatures, the potential minimum is initially located along the solid green line. As the temperature decreases, a smooth transition occurs, leading the minimum to shift to the blue line. This signifies the first step of the phase transition, which manifests as a second-order phase transition.
As the temperature continues to decrease, degenerate vacua emerge in both the $h$ and $\phi$ directions at a critical temperature $T_c$.  The global minima now corresponds to the red line. Subsequently, a significant second-step phase transition ensues, involving the tunneling of the system from the false vacuum to the true vacuum. This transition signifies a shift from the $Z_2$ broken phase to a phase where the $Z_2$ symmetry is restored, while the EW symmetry remains broken.
The nucleation of bubbles containing the true vacuum commences at a temperature denoted as $T_n$. Eventually, the phase transition reaches completion as the percolation temperature $T_p$ is reached.

\begin{figure}[htbp]
\centering
\includegraphics[width=0.465\textwidth]{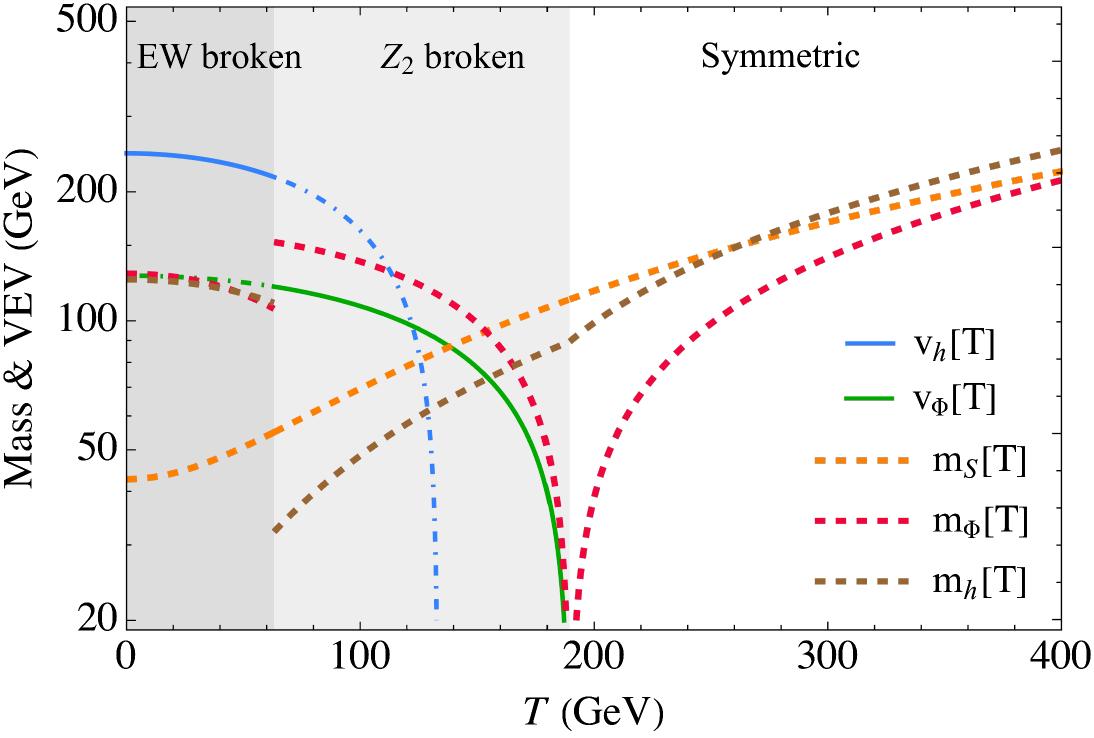}
\caption{Temperature dependent masses and VEVs. The thermal masses are depicted by dashed color curves and the VEVs are presented by the solid color lines.
The model parameters are taken from BP1 in Tab.~\ref{tab:PT-benchmarks}.}
\label{fig:mass-vevs}
\end{figure}

In such a thermal history, the masses of the scalar fields, specifically $\Phi$, $S$, and $H$, are temperature-dependent. 
%
During the symmetric phase, depicted in Fig. \ref{fig:mass-vevs}, the masses of $\Phi$, $S$, and $H$ decrease with temperature and are solely determined by $T^2$ dependence.
In the ``$Z_2$ broken'' phase, the particles' masses are governed by the VEV of $\Phi$ (green solid line). The mass of $\Phi$, denoted as $m_\Phi$, increases with the VEV $v_\Phi(T)$ (red dashed line). Meanwhile, due to the presence of small couplings, the mass variations of $S$ and $H$ are primarily driven by the $T^2$ term in Eqs. (\ref{eqn:ms-t}) and (\ref{eqn:mh-t}). 
In the ``EW broken'' phase, the VEV of $h$ (blue solid line) contributes to the masses of the scalar fields. As the system evolves, the masses of the scalar fields transition towards their zero-temperature values.

It is important to emphasize again that the thermal masses undergo changes with temperature across different phases, leading to temperature-dependent kinematics of DM production.
Therefore, the satisfaction of production thresholds cannot be guaranteed for all phases.

\section{FIMP DM productions during phase transitions}
\label{sec:FIMP}

The interactions between the FIMP DM and particles in the thermal bath, specifically the scalar field $\Phi$, are captured by the potential described by Eq. (\ref{eq:potential}). In this thermal history, multiple production processes arise, each governed by different couplings.
There are the exponential growth process $\Phi S \to SS$  and the three-body decay process $\Phi \to 3S$ driven by the coupling $\lambda_1$. Additionally, the $\Phi$-pair annihilation processes $\Phi \Phi \to SS$ (with couplings $\lambda_{S\Phi}$ and $\lambda_2$) and $\Phi \Phi \to \Phi S$ (with coupling $\lambda_2$) also contribute to production. 
Moreover, during the second phase where $v_\Phi$ is non-vanishing, a two-body decay production $\Phi \to 2S$ (governed by $\lambda_1$) occurs for a short period.
The accessibility of these production channels varies across different cosmological epochs, with some channels being kinematically closed in certain periods.

For simplicity, we firstly consider the coupled Boltzmann equations including the all production processes, without distinguishing between the different phases of the Universe. 
Further details regarding each production mechanism, including their respective kinematic properties and epoch-specific accessibility, are discussed in the following subsections.
The coupled Boltzmann equations for the evolution of both FIMP DM $S$ and WIMP-like particle $\Phi$ are given by, 
\begin{align}
\frac{d Y_{\Phi}}{d x} &= 
-\frac{2 \pi^2}{45} 
\frac{m_{\rm Pl} m_{\rm DM} \sqrt{g_{*}(x)}}{1.66 x^{2}}
\langle \sigma v \rangle_{\Phi\Phi\to hh}
\left( Y^{2}_{\Phi} - Y^{{\rm eq}2}_{\Phi} \right) \nonumber\\
&- \frac{3 m_{\rm Pl} x \sqrt{g_{*}(x)} }{1.66 m^{2}_{\rm DM} g_{s}(x)} 
\langle \Gamma_{\Phi\to3S} \rangle \left( Y_{\Phi} - Y^{3}_{S} \right) \nonumber\\
&-\frac{2 m_{{\rm Pl}}}{1.66 m^2_{\rm DM}} 
\frac{x \sqrt{g_{*}(x)}}{g_{s}(x)}
 \langle \Gamma_{\Phi 
\to SS} \rangle 
\left( Y^{{\rm eq}}_{\Phi} - Y^{2}_{S} \right)
\,,  
\end{align}

\begin{align}
\frac{d Y_{S}}{d x} &= 
-\frac{2 \pi^2}{45} \frac{m_{{\rm Pl}} m_{\rm DM}}{1.66 x^{2}}
\sqrt{g_{*}} \langle \sigma v \rangle_{\rm semi} \left( Y^{2}_{S} - Y^{{\rm eq}}_{\Phi} Y_{S} \right) \nonumber\\
&+\frac{4 \pi^2}{45} 
\frac{m_{\rm Pl} m_{\rm DM} \sqrt{g_{*}}}{1.66 x^{2}}
\langle \sigma v \rangle_{\Phi\Phi} 
\left( Y^{{\rm eq}2}_{\Phi} - Y^{2}_{S} \right) \nonumber\\
&+ \frac{3 m_{{\rm Pl}} x \sqrt{g_{*}(x)}}{1.66 m^{2}_{\rm DM} g_{s}(x)} 
\langle \Gamma_{\Phi\to 3S} \rangle \left( Y_{\Phi} - Y^{3}_{S} \right) \nonumber\\
&+\frac{2 m_{{\rm Pl}}}{1.66 m^2_{\rm DM}} 
\frac{x \sqrt{g_{*}(x)}}{g_{s}(x)}
 \langle \Gamma_{\Phi 
\to 2S} \rangle 
\left( Y^{{\rm eq}}_{\Phi} - Y^{2}_{S} \right) 
\,,
\label{eqn:cBE}
\end{align}
where the Planck mass $m_{\rm Pl} = 1.22 \times 10^{19}$ GeV. In this context, $m_{\rm DM}=m_S^0$ and $x=m_{\rm DM}/T$, while $g_{*}(x)$ and $g_{s}(x)$ represent the effective and entropic degrees of freedom of the Universe, respectively.
We define the yields $Y_{\Phi,S}\equiv n_{\Phi,S}/s$, where $n_{\Phi,S}$ corresponds to the number densities and $s$ represents the entropy density. 

The first equation governs the evolution of $\Phi$. On the right-hand side of the equation, the first term represents the annihilation of $\Phi$ into Higgs particles. The term $\langle \sigma v \rangle_{\Phi\Phi\to hh}$ denotes the thermal-averaged cross section multiplied by the relative velocity for the annihilation process. 
The second and third terms in the equation accounts for the decay of $\Phi$ into FIMP DMs.

The production of FIMP DM $S$ is described by the second equation.
Similarly, the first term on the right-hand side of the equation accounts for the contribution from the semi-production process, characterized by the thermal-averaged cross section $\langle \sigma v \rangle_{\rm semi}$.
The second term on the right-hand side represents the contribution from $\Phi$ pair annihilations. Here, $\langle \sigma v \rangle_{\Phi\Phi}$ denotes the thermal average of the annihilation cross section. 
The third and fourth terms correspond to the decay contribution of $\Phi$ into FIMP DM. Importantly, it should be noted that the decay process $\Phi\to SS$ only occurs during the second phase, owing to the non-zero VEV of the $\Phi$ particle.

The decay of the $\Phi$ particle into FIMP DM has a significant implication. It results in the eventual decay of the $\Phi$ particle before the onset of big bang nucleosynthesis (BBN).
Furthermore, even if the decay of $\Phi$ occurs after BBN, it does not contribute to the visible energy content. This property ensures that our model remains consistent with observational constraints arising from the abundances of light elements \cite{Kawasaki:2017bqm}.

The thermal average of the decay rate is defined as

\begin{equation}
\langle \Gamma_{\Phi\to 2S/3S} \rangle = \Gamma_{\Phi\to 2S/3S}  \frac{K_{1}(x)}{K_{2}(x)},
\end{equation}
where $K_{1,2}$ denote the modified Bessel functions of the second kind. 
The thermal average of cross section times velocity, $\langle \sigma v \rangle$, can be obtained by
\begin{align}
\langle \sigma v \rangle_{AB \to CD} &= 
\frac{1}{8 m_{A}^{2} m_{B}^{2} 
K_{2}\left(m_{A}/T\right) K_{2}\left(m_{B}/T\right)} \nonumber\\
&\times
\int_{\left(m_{A}+m_{B}\right)^{2}}^{\infty} d s \, 
\frac{\sigma_{AB \rightarrow C D}}{\sqrt{s}} p_{AB} 
K_{1}\left(\frac{\sqrt{s}}{T}\right) 
\,,
\end{align}
where $m_{A,B}$ are the masses of $A$ and $B$, $\sqrt{s}$ is the centre-of-mass energy, $p_{AB} = [s - ( m_{A} - m_{B} )^{2} ] [s - (m_{A} + m_{B} )^{2} ]$. 
In this study, the model is implemented by FeynRules \cite{Alloul:2013bka},  
then the decay rates and the scattering cross sections are calculated by FeynCalc \cite{Shtabovenko:2016sxi,Shtabovenko:2020gxv}. 
For the calculation, thermal masses are employed.

In order to facilitate a comprehensive analysis, we have carefully chosen specific parameter benchmarks to investigate and elucidate the distinct production mechanisms involved, including semi-production, three-body decay processes, and pair annihilation.
These benchmarks are outlined in Tab.~\ref{tab:DM}.

\begin{table}[htbp]
    \centering
    \begin{tabular}{|c|c|c|c|c|c|c|c|c|c|c|} 
  \hline
  Scenario &  $m_{S}^0 ({\rm GeV})$ &   $\lambda_{S\Phi}$ & $\lambda_1$&$\lambda_2$ \\ 
  \hline
   Semi-production & $42.9 $  & $10^{-19}$ & $10^{-7}$ & $10^{-19}$ \\
   \hline
   3-body decay & $42.7 $   & $10^{-19}$& $10^{-9}$& $10^{-19}$ \\
   \hline
   Pair-annihilation & $42.9 $ & ${6\times10^{-13}}$& $10^{-19}$ & $6\times10^{-13}$ \\
  \hline
 \end{tabular}  
    \caption{Benchmarks for the three different production scenarios.  $m^0_S $ represents the zero-temperature mass of FIMP $S$. The other parameters are taken from BP1 in Tab.~\ref{tab:PT-benchmarks}.}
    \label{tab:DM}
\end{table}

\subsection{Semi-production: an exponential growth process}
\label{subsec:semi}

We begin by focusing on the production dominated by the semi-production process ($\Phi \Phi \to \Phi S$). In this analysis, we consider the coupling $\lambda_1$ to be the dominant parameter, as in Tab.~\ref{tab:DM}, prioritizing its influence over other parameters.

The Boltzmann equation for the number density of FIMP can be written approximately as, 
\begin{equation}
\label{eq:BE-semi}
  \dot n_S+3H n_S \simeq \langle\sigma v\rangle_\mathrm{semi}\,n_S ~n_\Phi^\mathrm{eq}.
\end{equation}
In this equation, the term $n_S^2$ is neglected since it is considered to be negligible initially. 
In terms of the yield $Y_{S}$, the solution exhibits an exponential behaviour,
\begin{align}
Y_{S} = Y_S^{\rm initial} \exp \left[ 
\int_{x_{\rm init}}^{x} dx \, \frac{h_{\rm eff}(x) \langle \sigma v  \rangle_{\rm semi} 
n^{{\rm eq}}_{\Phi } }{x H(x)} 
\right]
\, ,
\end{align}
where $h_{\rm eff} (x)=1- \frac{d\log g_s(x)}{d \log x}$, $H(x)$ is the Hubble parameter. 
The assumption of equilibrium behavior is made for the distribution functions of DM. For a more accurate treatment of the semi-production processes, we refer the interested reader to the analysis presented in Ref. \cite{Bhatia:2023yux}. We present our results by numerically solving the complete Boltzmann Eq.~(\ref{eqn:cBE}) in the following.

\begin{center}
\begin{figure}[htbp]
\centering
\includegraphics[width=0.465\textwidth]{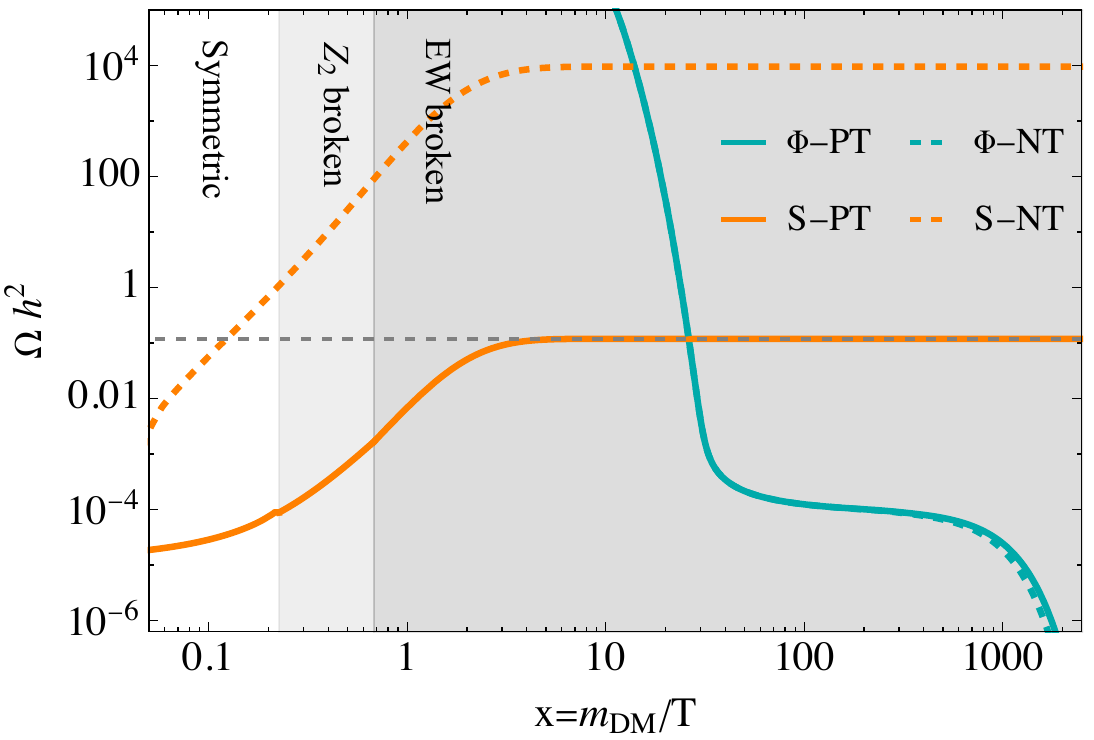}
\caption{Evolution of the DM relic density. Freeze-in production of FIMP $S$ via exponential growth is shown in orange solid line, i.e. from $\Phi S\to SS$. The cyan solid  line corresponds to the evolution curve of $\Phi$. 
The dashed curves stand for the evolution of $S$ and $\Phi$ without thermal phase transition effect (labeled by ``S-NT'' and ``$\Phi$-NT''). 
The gray dashed line indicates the correct relic density $\Omega h^2=0.12 \pm 0.001$. 
The parameters are taken as in Tab.~\ref{tab:DM}.}
\label{fig:semi-production}
\end{figure}
\end{center}

In Fig.~\ref{fig:semi-production}, we present the evolution of the DM relic density $\Omega h^2$ as a function of $x = m_{\rm DM}/T$, along with the evolution of the bath particle $\Phi$. The orange solid line corresponds to the FIMP DM production through the semi-production process $\Phi S\to S S$. Notably, we observe an exponential growth in the abundance, which persists until freeze-in at $x\simeq 3$, reaching the observed relic density of $\Omega h^2\simeq0.12$ \cite{Planck:2018vyg}.
Throughout all three phases, the FIMP abundance exhibits continuous and smooth growth.
We have examined the FIMP production without considering the thermal evolution history of the Universe. The orange dashed line represents the results obtained with the same set of parameters, demonstrating a different production pattern and significantly higher abundance. This discrepancy arises from the inclusion of the three-body decay process, facilitated by the condition $m^0_\Phi > 3 m^0_S$. Consequently, the production mode dominated by the decays of the bath particle surpasses the semi-production contribution.

Notice that the initiation of this process requires an initial population of DM to be present, which can originate from various sources, including remnants from the reheating process \cite{Takahashi:2007tz}, gravitational production effects \cite{Garny:2015sjg,Mambrini:2021zpp}, or ultraviolet freeze-in \cite{Moroi:1993mb,Bolz:2000fu}. 
For the sake of simplicity in our study, we assume an initial abundance of $Y^{\rm initial}_{S}=1.5 \times 10^{-15}$.

The cyan solid (dashed) line represents the evolution of the bath particle $\Phi$ with (without) thermal effects. 
$\Phi$ is in thermal equilibrium with the SM plasma, undergoing frequent annihilations with SM Higgs particles and vice versa.
In this particular scenario, $\Phi$ is transformed by $S$ to be a FIMP along its freeze-out trajectory. Subsequently, after EW symmetry breaking, $\Phi$ starts to decay into three FIMPs at a certain temperature, facilitated by the condition $m_\Phi>3m_S$, where the decay kinematics become energetically allowed. 
This is known as the superWIMP contribution \cite{Feng:2003xh}, appearing at $x \simeq 10^{4}$. However, its impact is negligible due to the significantly lower abundance of $\Phi$ at that time compared to the abundance of FIMPs.
Importantly, it is worth noting that the evolution without considering the thermal phase transitions closely resembles the thermal case. This implies that the behavior of WIMP-like particles is largely independent of temperature corrections and the thermal history.

\subsection{Three-body decay production}
\label{sec:3decay}

In this benchmark model, we set $m^0_\Phi> 3 m^0_S$  and a proper setting of other relevant couplings as listed in Tab.~\ref{tab:DM}. This choice of parameters allows us to investigate the specific scenario where the three-body decay channel plays a significant role in the production dynamics. Fig.~\ref{fig:mass-vevs} reveals that the mass relation between $\Phi$ and $S$ changes in time. When the mass of $\Phi$ is less than three times the mass of the FIMP DM, the three-body decay channel becomes kinematically forbidden. Therefore, in this case, we have a complete different DM evolution. 
The  decay rate of $\Phi\to SSS$ is given by \cite{Costa:2022oaa}, 
\begin{align}\label{eqn:3body-decay}
\Gamma = \frac{m_\Phi}{256 \pi^3} 
|\mathcal{M}|^2 
\int^{1-3a}_{2 \sqrt{a}} dy \sqrt{\frac{(y^2-4a)(1-3a-y)}{1+a-y}}
\,,
\end{align}
with $a=m^2_S/m^2_\Phi$, and $|\mathcal{M}|^2=36\lambda_1^2$.


\begin{center}
\begin{figure}[htbp]
\centering
\includegraphics[width=0.465\textwidth]{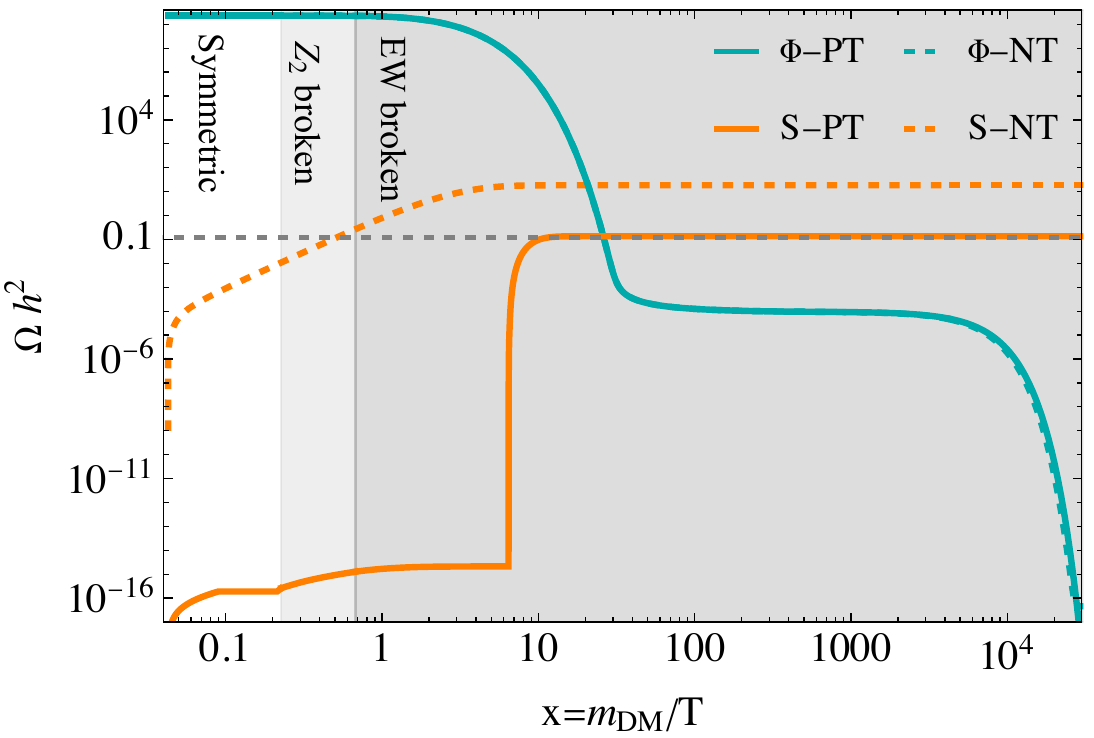}
\caption{Evolution of the DM relic density. FIMP production via the three body decay process, ie. $\Phi\to SSS$. The solid color lines stand for the densities that take into account the thermal effect. The dashed curves stand for the evolution of $S$ and $\Phi$ without thermal phase transition effect (labeled by ``S-NT'' and ``$\Phi$-NT''). The parameters are taken from Tab.~\ref{tab:DM}. }
\label{fig:3decay}
\end{figure}
\end{center}

Neglecting the thermal effects, the orange-dashed line in Fig.~\ref{fig:3decay} represents the freeze-in production, characterized by a gradual increase in the yield until freeze-in occurs at $x\sim3$. This behavior is consistent with the typical features of the freeze-in mechanism. 
In contrast, when thermal effects are incorporated into the Boltzmann equation, the evolution of DM undergoes significant modifications. The orange-solid line in Fig.~\ref{fig:mass-vevs} demonstrates that the DM abundance experiences only marginal accumulation during the symmetric and ``$Z_2$ broken'' phases, primarily driven by the annihilation of $\Phi$ pairs. However, a notable shift occurs during the EW broken phase around $x\sim10$, coinciding with the opening of the $\Phi\to SSS$ decay channel. This leads to a rapid growth in DM abundance. 
Notably, the final abundances obtained from the two approaches, employing the same parameter set, exhibit significant deviations. Consequently, we conclude that thermal corrections play a pivotal role and cannot be overlooked.

The cyan lines represent the evolution of the WIMP-like particle $\Phi$, which exhibits a similar evolution pattern to the semi-production scenario.

\subsection{Production from $\Phi$-pair annihilation}

The third production scenario involves the dominant production of FIMPs through $\Phi$ pair annihilations, specifically the processes $\Phi\Phi\to SS$ and $\Phi\Phi\to\Phi S$. To investigate these processes, we utilize the benchmark parameters outlined in Tab.~\ref{tab:DM}. By increasing the values of $\lambda_{S\Phi}$ and $\lambda_2$ (which we assume to be equal for simplicity), the annihilation production becomes the prevailing mechanism.

\begin{center}
\begin{figure}[!htbp]
\centering
\includegraphics[width=0.465\textwidth]{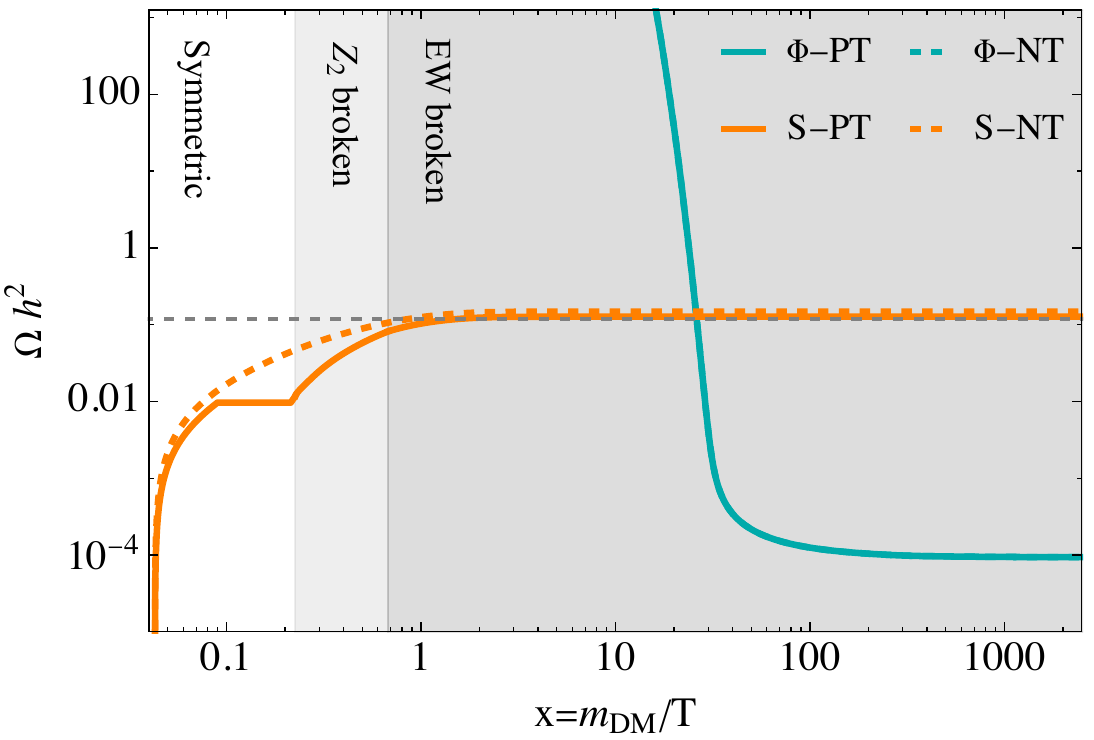}
\caption{
Evolution of the DM relic density. FIMPs from $\Phi$-pair annihilation production, i.e. $\Phi\Phi\to SS$ and  $\Phi\Phi\to \Phi S$. 
The solid color lines stand for the densities that take into account the thermal effect. The dashed curves stand for the evolution of $S$ and $\Phi$ without thermal phase transition effect (labeled by ``S-NT'' and ``$\Phi$-NT''). The parameters are taken from Tab.~\ref{tab:DM}. 
}
\label{fig:pair}
\end{figure}
\end{center}

In Fig. \ref{fig:pair}, we present the production curves for this scenario, showcasing the evolution of both the FIMP DM and the bath particle $\Phi$. 
The solid lines represent the evolution considering the complete thermal history, while the dashed lines illustrate cases where thermal masses are not taken into account. 
Discernible discrepancies are observed in the FIMP DM patterns between the solid orange curve and the dashed curve. The solid orange curve exhibits a typical two-step pattern \cite{Bian:2018bxr}, which serves as the primary distinguishing feature from the non-thermal scenario.
Furthermore, the cyan lines (both solid and dashed) exemplify the thermal freeze-out of $\Phi$ around $x\sim20$. However, based on the chosen model parameters, the relic abundance arising from $\Phi$ can be safely disregarded.

\section{Predictions on Gravitational Waves}
\label{sec:GWs}

In addition to the diverse DM production mechanisms in this model, the phase transitions are also accompanied by the generation of GWs.

An increasing number of studies \cite{Alves:2018jsw,Guo:2021qcq,Wang:2020jrd,Zhou:2022mlz,Xie:2020bkl,Bian:2019szo,Caldwell:2022qsj} on the GW signals from the cosmological FOPT show that three mechanisms (bubble collisions, sound waves, and turbulence) are capable of producing tensor perturbations. To analyze the GW spectrum emitted during the FOPT, we should calculate the phase transition parameter $\alpha$, and the phase transition duration parameter $\beta$, which are given by, 
\begin{align}
&\alpha = \frac{\Delta \rho}{\rho_R} \;,\\
&\frac{\beta}{H_*} = T \frac{d(S_3/T)}{d T}|_{T = T_*} , 
\end{align}
where $H_*$ is the Hubble constant at the characteristic  temperature $T_*$. $\rho_R$ is the radiation energy of the bath which can be written as $\rho_R = \pi^2 g_* T^4/30$, and $\Delta_\rho$ is the released latent heat during the phase transition, which is given by $\Delta_\rho = \rho^{ h-{\rm vac}}(T_*) - \rho^{\rm \Phi-vac}(T_*)$. The energy density is \cite{Enqvist:1991xw} \begin{align}
\rho^{h(\Phi)-{\rm vac}}(T) = V_T(v_{h(\Phi)},T) - T \frac{\partial V_T( v_{h(\Phi)},T)}{\partial T } \;.
\end{align}
In this study, we assume $T_* = T_p$ when  calculate both $\alpha$ and $\beta/H_*$. 

\begin{figure}[htbp]
\centering
\includegraphics[width=0.465\textwidth]{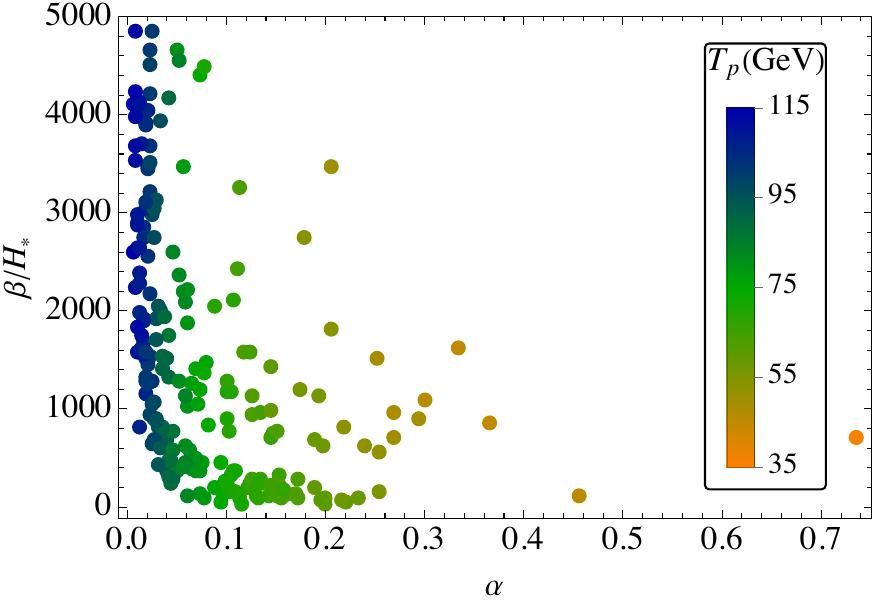}
\caption{The relation among key parameters of the phase transition $\beta/H_*$, $\alpha$ and $T_p$. The scatter-points with a lower percolation temperature $T_p$ has the higher energy budge $\alpha$ and smaller $\beta/H_*$. }
\label{fig:alphabeta}
\end{figure}

In Fig. \ref{fig:alphabeta}, we present the relationship between $\alpha$ and $\beta/H_*$ as a function of the percolation temperature. It is evident that a lower percolation temperature $T_p$ corresponds to a higher energy budget $\alpha$ and a smaller value of $\beta/H_* $. This indicates a greater potential for generating detectable GW signals. 
One of the selection principles of the BPs in Tab.~\ref{tab:PT-benchmarks} is based on this observation. Further details regarding the dependence on other model parameters are provided in the Appendix \ref{apdx:alphabeta}.

We adopt the approach outlined in Refs. \cite{Wang:2020jrd,Zhou:2022mlz}, taking three commonly coexisting sources into consideration, and the whole energy spectrum can be written as. 
\begin{align}
\Omega_{\rm GW} h^2 = \Omega_{\rm col} h^2 + \Omega_{\rm sw} h^2 + \Omega_{\rm turb} h^2.
\end{align}
Specific equations of GW spectrum can be found in Appendix\ref{appd-gw-spe}.

\begin{figure}[htbp]
\centering
\includegraphics[width=0.465\textwidth]{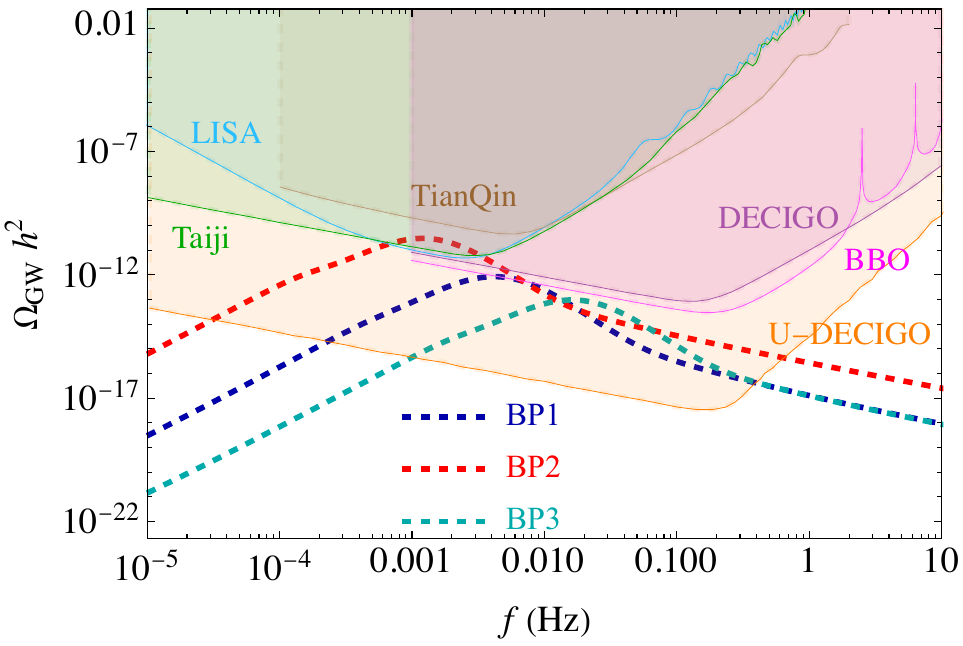}
\caption{The GW signals spectrum for the FOPT. The parameters of  three BPs can be found in Tab.~\ref{tab:PT-benchmarks}. The solid color curves stand for the sensitivities of the projected detectors.}
\label{fig:gw}
\end{figure}

Fig.~\ref{fig:gw} shows the GW signals for the three BPs outlined in Tab.~\ref{tab:PT-benchmarks}, along with the sensitivity curves of the future GW experiments. These include space-based laser interferometers such as LISA \cite{LISA:2017pwj}, Taiji \cite{TianQin:2015yph, Hu:2017yoc, TianQin:2020hid}, TianQin \cite{Hu:2017mde, Ruan:2018tsw}, BBO \cite{Crowder:2005nr}, and DECIGO \cite{Seto:2001qf, Kawamura:2020pcg, Kudoh:2005as}.
As observed in Fig.~\ref{fig:gw}, the GW signals generated from the FOPT of the three BPs exhibit peak frequencies in the range of $10^{-3} \sim 10^{-1}$ Hz, with peak amplitudes of approximately $10^{-10}$, $10^{-12}$, and $10^{-14}$ for the three respective BPs. These predicted GW signals are within the sensitivities of the projected detectors, including LISA, Taiji, DECIGO, BBO, and U-DECIGO.

\section{Conclusions}
\label{sec:conclusion}

In this study, we have investigated the phase-dependent production of FIMP DM within a $Z_2$ model incorporating two $Z_2$-odd scalar fields: the portal particle $\Phi$ and the DM $S$.
Our primary focus lies on the potential occurrence of a two-step phase transition pattern, which gives rise to  the thermal variations in the masses of $\Phi$ and $S$, resulting in a distinctive evolution of FIMP production.
Through our analysis, we have identified dominant mechanisms responsible for generating the FIMP $S$. These mechanisms include the semi-production process, as well as traditional freeze-in productions from pair annihilations of parent particles and a three-body decay process.
By comparing the freeze-in productions with and without considering the finite temperature effects, our results emphasize the significance of incorporating the thermal history. 
Furthermore, the occurrence of a strong two-step FOPT is accompanied by the generation of GW signals, presenting promising prospects for future detection using interferometers.
Our study provides valuable insights into the intricate interplay between the thermal history, DM production mechanisms, and the resulting signatures of GWs.

\section*{Acknowledgments}
We thank Bin Zhu, Zhi-Long Han and Ligong Bian for helpful discussions. 
The work by X.L. was supported by the National Natural Science Foundation of China under Grant Nos. 12005180, 12275232, by the Natural Science Foundation of Shandong Province under Grant No. ZR2020QA083, and by the Project of Shandong Province Higher Educational Science and Technology Program under Grants No. 2022KJ271.
W.C. was supported by Chongqing Natural Science Foundation project under Grant No. CSTB2022NSCQ-MSX0432, by Science and Technology Research Project of Chongqing Education Commission under Grant No. KJQN202200621, and by Chongqing Human Resources
and Social Security Administration Program under Grants No. D63012022005.
R.Z. was supported by supported by the National Natural Science Foundation of China under Grant No. 12305109, Chongqing Natural Science Foundation project under Grant No. CSTB2022NSCQ-MSX0534, and by Science and Technology Research Project of Chongqing Municipal Education Commission under Grant No.KJQN202300614.

\section{Appendix}

\subsection{Thermal masses}
\label{apdx:thermalmass}

The thermal masses of the scalars ($m_{S,\Phi,h}(T)$) are different in the symmetric phase, the $Z_2$ broken phase, and the EW symmetry broken phase. 
In the symmetric phase, the field dependent thermal masses can be obtained by 
\begin{eqnarray}
m_S^{sys}(T)&=&m_S(T)|_{\langle h, \Phi\rangle \to 0}\;,\\
m_\Phi^{sys}(T)&=&m_\Phi(T)|_{\langle h, \Phi\rangle\to 0}\;,\\
m_h^{sys}(T)&=&m_h(T)|_{\langle h, \Phi\rangle \to 0}\;.
\end{eqnarray}
In the $Z_2$ broken phase, we have
\begin{eqnarray}
m_S^{\slashed{Z_2}}(T)&=&m_S(T)|_{\langle h\rangle \to 0, \langle \Phi\rangle \to v_\Phi(T)  }\;,\\
m_\Phi^{\slashed{Z_2}}(T)&=&m_\Phi(T)|_{\langle h\rangle \to 0, \langle \Phi\rangle \to v_\Phi(T) }\;,\\
m_h^{\slashed{Z_2}}(T)&=&m_h(T)|_{\langle h\rangle \to 0, \langle \Phi\rangle \to v_\Phi(T)  }\;. 
\end{eqnarray}
In the EW broken phase, we have,
\begin{eqnarray}
m_S^{\slashed{EW}}(T)&=&m_S(T)|_{\langle h\rangle \to v_h(T), \langle \Phi\rangle \to 0}\;,\\
m_\Phi^{\slashed{ EW}}(T)&=&m_\Phi(T)|_{\langle h\rangle \to v_h(T), \langle \Phi \rangle \to 0 }\;,\\
m_h^{\slashed{EW}}(T)&=&m_h(T)|_{\langle h\rangle \to v_h(T), \langle \Phi\rangle \to 0}\;.
\end{eqnarray}
The thermal masses corresponding to each phase will be utilized in the numerical calculations.

\subsection{Dependence of $\beta/H_*$, $\alpha$ on model parameters}
\label{apdx:alphabeta}

In this subsection, we analyze the dependence of the key parameters, namely $\alpha$ and $\beta/H_*$, of the FOPT on the model parameters in scatter plots.

\begin{figure}[htbp]
\centering
\includegraphics[width=0.465\textwidth]{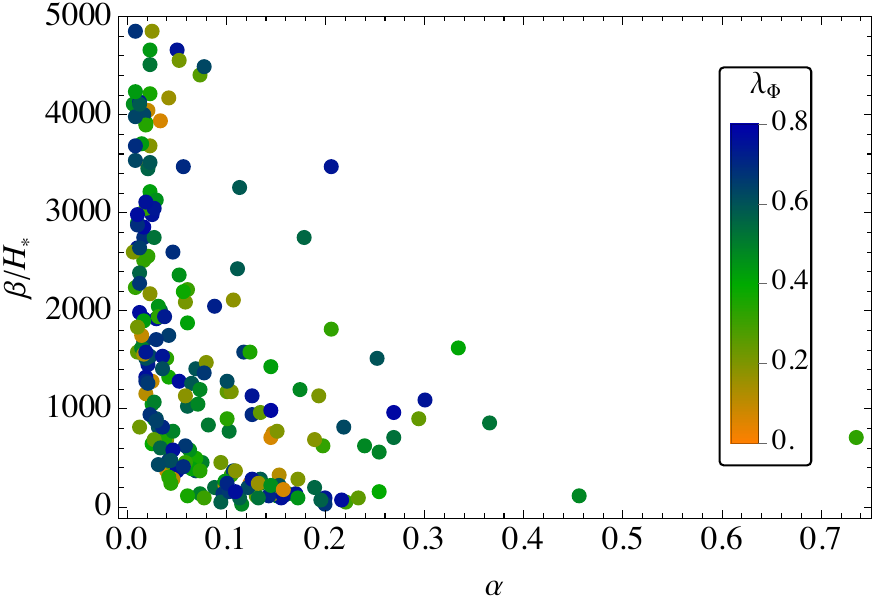}
\caption{The relation among key parameters of the phase transition $\beta/H_*$, $\alpha$ and coupling $\lambda_{\Phi}$. The self-coupling  $\lambda_{\Phi}$ for FOPT points typically ranges around [0, 0.8], and the majority of FOPT points have a higher self-coupling. }
\label{fig:alphabetalamphi}
\end{figure}

Figure \ref{fig:alphabetalamphi} showcases the distribution of these key parameters. Specifically, it demonstrates the variations of $\beta/H_*$ and $\alpha$ as a function of the coupling parameter $\lambda_{\Phi}$. We utilize color-changing points to represent different values of the coupling parameter $\lambda_{\Phi}$.  The range of the self-coupling parameter $\lambda_{\Phi}$ is typically within [0, 0.8] to ensure the occurrence of the FOPT. Notably, the majority of points exhibit higher values of the $\lambda_\Phi$.

\begin{figure}[htbp]
\centering
\includegraphics[width=0.465\textwidth]{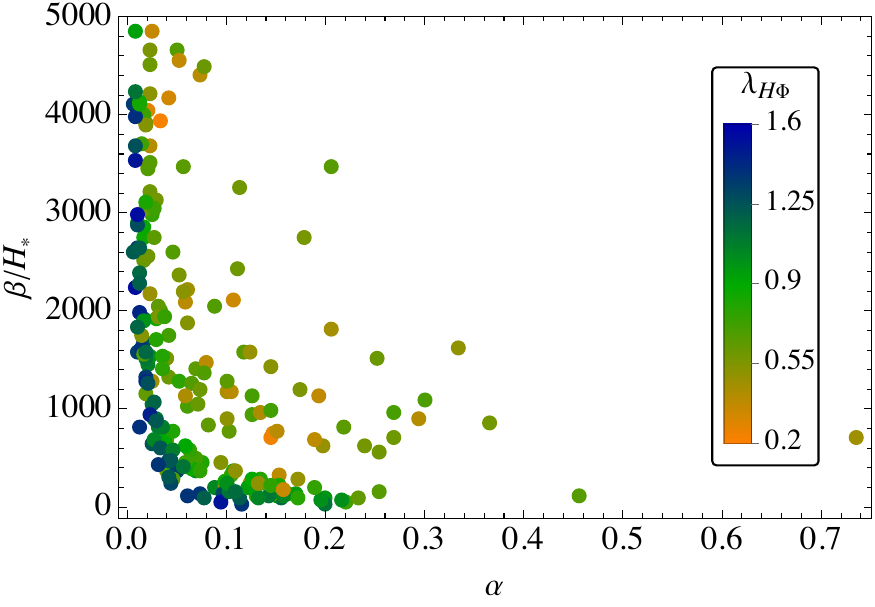}
\caption{The relation among key parameters of phase transition $\beta/H_*$, $\alpha$ and coupling $\lambda_{H\Phi}$.}
\label{fig:alphabetlamhphi}
\end{figure}

Fig.~\ref{fig:alphabetlamhphi} depicts the relationship between $\beta/H_*$, $\alpha$, and the coupling $\lambda_{H\Phi}$. Notably, the coupling $\lambda_{H\Phi}$ has a more pronounced impact on the FOPT compared to the self-coupling parameter $\lambda_{\Phi}$.
In the plot, we observe that the majority of FOPT points with lower values of $\alpha$ are concentrated within the range of $\lambda_{H\Phi}\in$[0.4, 1]. This suggests that the coupling $\lambda_{H\Phi}$ plays a crucial role in influencing the FOPT dynamics.

\begin{figure}[htbp]
\centering
\includegraphics[width=0.465\textwidth]{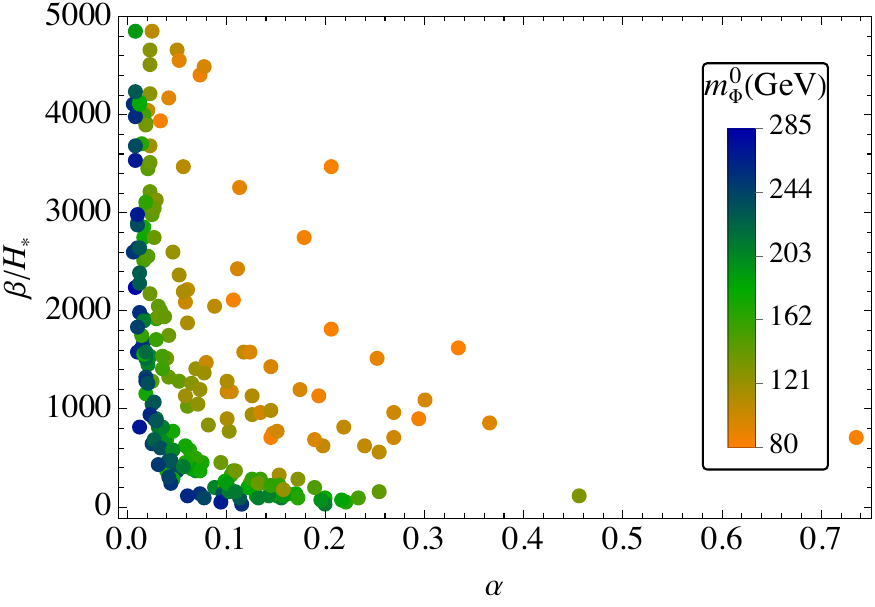}
\caption{The relation among key parameters of phase transition $\beta/H_*$, $\alpha$ and $m^0_\Phi$. The distribution trend of $m_\Phi$ is similar to that of $\lambda_{H\Phi}$.}
\label{fig:alphabetamphi}
\end{figure}

Fig.~\ref{fig:alphabetamphi} illustrates  the relationship between the key parameters and $m^0_\Phi$, where $m^0_\Phi$ is linearly related to the coupling $\lambda_{H\Phi}$. As a result, the distribution trend of $m_\Phi$ exhibits similarity to that of $\lambda_{H\Phi}$.

\subsection{Three sources of GW energy spectrum }
\label{appd-gw-spe}
We utilize the widely-accepted envelope approximation to calculate gravitational waves from bubble collisions \cite{Huber:2008hg},  
\begin{align}\label{eq:Omegaenv}
\Omega _{\rm coll} h^2(f)=&1.67\times10^{-5}\left(\frac{100}{g_{*}}\right)^\frac13
\left(\frac{\beta}{H_*}\right)^{-2}\left(\frac{\kappa_\Phi\alpha}{1+\alpha}\right)^2 \nonumber\\
&\times \frac{0.11v_w^3}{0.42+v_w^2}
\frac{3.8\left(f/f_{\rm coll}\right)^{2.8}}{1+2.8\left(f/f_{\rm coll}\right)^{3.8}},
\end{align}   
where $v_w$ is the bubble wall velocity, and $k_\Phi$ is the efficient factor which represents the proportion of the latent heat deposited within a thin shell. The peak frequency is located at  
\begin{align}\label{eq:fenv}
f_{\rm coll}=&1.65\times10^{-5}\left(\frac{g_*}{100}\right)^\frac16\frac{T_\star}{100\mathrm{GeV}} \frac{\beta}{H_*} \nonumber \\
&\times \frac{0.62}{1.8-0.1v_w+v_w^2}\,\mathrm{Hz} .
\end{align}

According to the sound shell model with taking the lifetime suppression factor into account, the sound wave contribution \cite{Hindmarsh:2013xza,Hindmarsh:2015qta} $\Omega _{\rm sw} h^2$ and the peak frequency $f_{\rm sw}$ can be written as
\begin{align}
\label{eq:sw}
\Omega _{\rm sw} h^2 (f)&=1.64 \times 10^{-6}(H_*\tau_{sw})\left(\frac{\beta}{H_*}\right)^{-1}\nonumber\\
&\times\left(\frac{\kappa \alpha }{1+\alpha }\right)^2
\left(\frac{g_*}{100}\right)^{-\frac{1}{3}} v_w(8\pi)^{1/3}
\left(\frac{f}{f_{\rm sw}}\right)^3\nonumber\\
&\times 
\left(\frac{7}{4+3 \left(f/f_{\rm sw}\right)^2}\right)^{7/2}\;,
\end{align}
with
\begin{align}
f_{\textrm{sw}}&=1.9\times10^{-5}\frac{1}{v_{w}}\frac{\beta}{H_*} \left( \frac{T_{\ast}}{100\textrm{GeV}} \right) \left( \frac{g_{\ast}}{100}\right)^{1/6} \textrm{Hz} \;,
\end{align}
and $\tau_{sw}$=${\rm min}\left(\frac{1}{H_*},\frac{R_*}{\bar{U}_f}\right)$, $H_*R_*$=$v_w(8\pi)^{1/3}(\beta/H)^{-1}$. Here, $\bar{U}_f$ is the root-mean-square fluid velocity that can be approximated as\,\cite{Hindmarsh:2017gnf,Caprini:2019egz,Ellis:2019oqb,Guo:2020grp}
\begin{equation}
\bar{U}_f^2\approx\frac{3}{4}\frac{\kappa\alpha}{1+\alpha}\;,
\end{equation}
$\kappa$ is the fraction of the released energy into the kinetic energy of the plasma, which can be calculated given ~\cite{Espinosa:2010hh}.
\begin{equation}
\kappa = \frac{\sqrt{\alpha}}{0.135+\sqrt{0.98+\alpha}}
\end{equation}

There is another source of GW power spectrum generated by turbulence mechanism. The contribution to the GW power specturm $\Omega h_{\textrm{turb}}^{2}$ and the peak frequency are given by \cite{Caprini:2009yp,Binetruy:2012ze}
\begin{align}
\Omega _{\textrm{turb}} h^{2}=&\,3.35\times10^{-4}\left( \frac{\beta}{H_*}\right)^{-1} \left(\frac{\kappa_{\text{turb}}
\alpha}{1+\alpha} \right)^{3/2} \left( \frac{100}{g_{\ast}}\right)^{1/3}\nonumber\\
&\,\times v_{w} \cdot \frac{(f/f_{\textrm{turb}})^{3}}{[1+(f/f_{\textrm{turb}})]^{11/3}(1+8\pi f/h_{\ast})} \;, \\
f_{\textrm{turb}}=&2.7\times10^{-5}\frac{1}{v_{w}}\frac{\beta}{H_*}  \left( \frac{T_{\ast}}{100\textrm{GeV}} \right) \left( \frac{g_{\ast}}{100}\right)^{1/6} \textrm{Hz} .
\label{eq:mhd}
\end{align}
where the factor $\kappa_{\text{turb}}$ is the fraction of energy transferred to the MHD turbulence and can be roughly estimated as $\kappa_{\text{turb}} \approx \epsilon \kappa$ with $\epsilon\approx$ 5 $\sim$ 10\%~\cite{Hindmarsh:2015qta}. 
And we set $\epsilon\approx 0.1$ in this study.

\bibliography{biblio}

\end{document}